\documentclass[12pt]{iopart}
\usepackage[dvips]{graphicx}
\usepackage{iopams} 

\def\real{\mathbb{R}}
\def\Tr{\mathop{\rm Tr}\nolimits}
\def\argmax{\mathop{\rm argmax}}

\newtheorem{lem}{Lemma}
\newtheorem{thm}{Theorem}

\begin{document}
\title[General theory for decoy-state quantum key distribution]{General theory for decoy-state quantum key distribution with arbitrary number of intensities}
\author{Masahito Hayashi}
\address{ERATO-SORST Quantum Computation and Information Project, JST\\
5-28-3, Hongo, Bunkyo-ku, Tokyo, 113-0033, Japan}
\ead{masahito@qci.jst.go.jp}
\begin{abstract}
We develop a general theory for quantum key distribution (QKD)
in both the forward error correction and the reverse error correction cases 
when the QKD system is equipped with phase-randomized coherent light with arbitrary number of decoy intensities. 
For this purpose, 
generalizing Wang's expansion,
we derive a convex expansion of the phase-randomized coherent state. 
We also numerically check that the asymptotic 
key generation rates are almost saturated 
when the number of decoy intensities is three.
\end{abstract}

\pacs{03.67.Dd,03.67.Hk,03.67.-a,05.30.Jp}

\vspace{2pc}
\noindent{\it Keywords}: 
decoy method, phase-randomized coherent state,
key generation rate, arbitrary number of intensities

\submitto{\NJP}
\maketitle

\section{Introduction}
The BB84 protocol proposed by Bennett and Brassard\cite{bene} has been
known as a famous protocol guaranteeing information theoretical security. 
Its security has been proved by Mayers\cite{mayer1}
in the single-photon case. However, implemented channels usually have
imperfect quantum state generators, which generate not a perfect
single-photon but a mixture of the vacuum state, the single-photon
state, and the multi-photon state. 
In fact, many implemented quantum
communication systems are equipped with phase-randomized weak coherent light,
which can be regarded as the above type of mixed state\cite{SGGRZ,KNHTKN04,GYS,TMTT05,Y-S}.
In order to guarantee the security in the above imperfect 
quantum communication systems, 
Gottesman et al. (GLLP)\cite{GLLP}, and Inamori et al. (ILM)\cite{ILM}
have obtained a useful formula for the asymptotic key generation (AKG) rate,
which needs the counting rate and the phase error rate of single photon state.
If we treat the worst case of these parameters,
the obtained AKG rate become worse.
In order to estimate these parameters properly,
Hwang \cite{hwang} proposed the decoy method,
in which we randomly choose different intensities of the coherent light. 
In the decoy method, a pulse for final key generation is called the signal pulse
while a pulse only for estimation of these parameters is called 
a decoy pulse.
When $k$ intensities are choosed for decoy pulses, 
we need to prepare $k+1$ intensities.
Therefore, as was pointed out by Lo et al.\cite{LMC},
for the estimation of the counting rate of single photon state,
we have to treat an infinite number of the unknown parameters 
only with $k+1$ equations
even if the counting rates of individual number states 
do not depend on the basis.
(Note that the counting rates for 
multi-photon states depend on the basis as well as
the number of photon.
However, we consider this special case for simplicity in introduction.)
In this case, its estimate can be derived 
from the restriction for all unknown parameters.
Hence, lager number of intensities yields a more precise estimate of 
the counting rate of single photon state.
The phase error rate of single photon state has the same characterization.

In order to reduce the cost of its realization, Wang \cite{wang} proposed to choose the minimum decoy intensity to be the vacuum.
By using this idea, QKD system can be implemented with $k$ kinds of intensities essentially.
In the following, we adopt this assumption.
In the case of $k=2$, Wang \cite{wang} also proposed
an expansion of the phase-randomized coherent states
$\sum_{n=0}^{\infty} \frac{e^{-\mu}\mu^n}{n!}|n \rangle \langle n|$,
which describe the signal pulse and the decoy pulses.
His expansion is different from the conventional expansion by number states
in that the signal and decoy states can be described by stochastic mixtures of 
at most four states.
Under his expansion, we obtain three constraint equations with four unknown parameters.
He derived an estimate of the counting rate of the single photon state.
Ma et al.\cite{XBZL} also derived the same estimate by a different method.
Ma et al.\cite{XBZL} and Wang \cite{wang2} obtained an estimate of the phase error rate of the single photon state, and completed the estimation of 
AKG rate in the case of $k=2$, independently.
Ma et al.\cite{XBZL} pointed out that if the second smallest intensity goes to $0$,
the required parameters can be solved perfectly.
So, we obtain the best AKG rate, which is called the {\it universal upper bound}\footnote{Indeed, if we use a good code instead of random coding, there is a possibility to improve this bound.
However, if we use random coding it is impossible to improve the 
universal upper bound.}.
However, as was mentioned by them, it is impossible to control the infinitesimal intensity.
Further, even if it is possible, if we choose small intensity,
we cannot estimate the required parameters properly because the infinitesimal intensity makes our estimate fragile for statistical fluctuation.
Thus, for a precise estimate,
we need to fix the minimum difference between two intensities.
That is, it is suitable to compare the qualities of our estimates 
by changing the number of decoy intensities under this constraint.
For a good survey, please see Wang \cite{wang3}.

Now, we trun to the improvement of AKG rate.
After GLLP-ILM's result, Lo\cite{Lo}
conjectured an improvement of their AKG rate
with a rough idea of its proof. Based on his conjecture,
Boileau-Batuwantudawe-Laflamme (BBL)\cite{BBL} discussed a further
improvement of the rate by taking into account the effects of dark
counts in the detector. They pointed out that the AKG rate in the
forward error correction case is different with respect to that in the
reverse case. In the following, these rates will be called BBL 
formulas.
Lo's conjecture has been proved by Hayashi \cite{H1} and Koashi
\cite{Koashi}, independently. 
Hence, it is suitable to apply 
these formulas to the decoy method.

In this paper, we treat the case of arbitrary number $k$ of decoy intensities
with BBL formulas
when the intensity can be controlled\footnote{
For the case when the intensity fluctuates,
see the papers \cite{Wang4,Wang5} 
and the back note [26] of the paper \cite{HHHT}.}.
For this purpose, we generalize Wang's expansion of 
the phase-randomized coherent states,
in which $k+1$ phase-randomized coherent states
are given by convex combinations of $k+2$ states,
which are called {\it basis states}.
The idea of this generalization is trivial, however, 
it is not trivial to derive the concrete forms of
$k+2$ basis states and to check their positivity.
It is also needed to give the AKG rate formula 
by using the counting rates and the phase error rates
of each intensities based on this expansion.
For this purpose, 
we generalize mean value theorem and the concept of ``difference''.
In the conventional mean value theorem,
we treat the derivative of a given function $f$ and
the difference between outcomes of the function $f$ in two points.
In this paper, we extend the concept of ``difference''
to the case of $n+1$ points,
and derive its formula based on the $n$-th derivative.
We also derive a formula for the generalized  ``difference''.
Using these formulas, we give
 concrete forms of $k+2$ basis 
states and showed their positivity.
We also derive a formula for estimating 
the counting rate and the phase error rate of the single photon state
from the counting rates and the phase error rates of signal and decoy state.
Then, the AKG rate formulas with BBL formulas are obtained.
Moreover, we numerically compare 
the AKG rates of the cases of $k=2,3,4$ and the universal upper bound
in a proper model.
As our result, 
the AKG rates of the case of $k=3$ attain the universal upper bound
within 1\% error in casein the forward and reverse error case.
That is, further great improvement cannot be expected even if
the number $k$ increases more than $3$.

The organization of this paper is given as follows. Section \ref{s2}
gives the AKG rates in both the forward and reverse cases as
conjectured by BBL\cite{BBL}, taking into account the effect of dark
counts. 
In Section \ref{s3}, we treat the universal upper bound in a
proper channel model.
In Section \ref{s4}, we review the results by Wang \cite{wang,wang2}, 
Ma et al.\cite{XBZL}.
In Section \ref{s5}, we generalize Wang's expansion of 
the phase-randomized coherent state
$\sum_{n=0}^{\infty} \frac{e^{-\mu}\mu^n}{n!}|n \rangle \langle n|$.
In Section \ref{s6}, we give a lower bound of AKG rate
based on the counting rates and phase error rates of respective intensities. 
In Section \ref{s7}, we
numerically compare the AKG rates in the case of $k=2,3,4$
and its universal upper bound.
In Section \ref{s8}, 
the generalized  ``difference'' is introduced,
and all theorems are proved.

\section{AKG rates with dark counts effect}\label{s2}
When the transmitted pulse is given as a mixture of the vacuum, single-photon,
and multi-photon, 
taking into account the effect of dark counts in the detector,
BBL\cite{BBL} conjectured the following AKG rate
by use of Lo\cite{Lo}'s conjecture,
which is proved by 
Koashi \cite{Koashi} and Hayashi \cite{H1} 
with the threshold detector independently.
Further, Hayashi \cite{H1} allows loss of detector
if it is independent of the measurement basis.
That is, we assume the same receiver model as Hayashi \cite{H1}.
Let $q^0$ and $q^1$ be the probabilities
detecting the the vacuum state and the single-photon state
except for dark counts in detector.
Let $\mu$ be the intensity of the signal pulse, which produces the raw keys.
Then, We denote the dark count rate in the detector, 
the phase error rate concerning the single-photon state
counted except for dark counts, 
the bit error probability of signal pulse,
and the counting rate of signal pulse
by $p_D$, $r^1$, $s_\mu$, and $p_{\mu}$, respectively.
When we use the phase-randomized coherent light with the intensity $\mu$,
$e^{-\mu}$ and $\mu e^{-\mu}$ are the probabilities generating
the vacuum state and the single-photon state.
In this case, Eve has no information concerning Bob's bit of the 
pulses detected by dark counts and 
Alice's bit of the pulses whose state is the vacuum state.
Hence, the AKG rate with the forward error correction
is different from that with the reverse error correction,
and they are given as follows. (See BBL\cite{BBL} or Section IV A of Hayashi\cite{H1}.)
\begin{eqnarray}
I_{\to}:=
&\frac{1}{2}
\left(
\mu e^{-\mu}
 q^1 (1-\overline{h}(r^1))
+e^{-\mu}(q^0+p_D)
- p_{\mu}\eta(s_\mu)
\right)
\label{2-16-6}\\
I_{\leftarrow}
:=&\frac{1}{2}\left(
\mu e^{-\mu} q^1 (1-\overline{h}(r^1))
+p_D- p_{\mu}\eta(s_\mu)
\right),
\label{2-16-7}
\end{eqnarray}
where $\eta(s)$ is chosen so that 
$1-\eta(s)$ is 
the coding rate of the classical error correction code
and 
$\overline{h}(x)$ is given by
\begin{eqnarray*}
\overline{h}(x)=
\left\{
\begin{array}{ll}
-x\log_2 x-(1-x)\log_2(1-x) & \hbox{ if } x \le \frac{1}{2} \\
1& \hbox{ if } x > \frac{1}{2} .
\end{array}
\right.
\end{eqnarray*}
In the above formula, we regard the simultaneous event of a dark count and a normal count
as a dark count.
This is because the collision of both photons disturbs the information of the normal count.

These are the rates per one pulse sent by Alice.
Thus, the coefficient $\frac{1}{2}$ corresponds to 
the probability that the basis of Alice coincides with that of Bob.
For protocols attaining these rates,
see Hayashi \cite{H1}.

The dark count rate in the detector should be measured before the sending 
the quantum communication.
In the above formulas, 
the parameters $q^0,q^1$ and $r^1$ are not known {\it a priori}.
These cannot be estimated 
if only the pulse with the intensity $\mu$ is transmitted.
In the decoy method \cite{hwang}, we randomly change the intensity $\mu$ and estimate these parameters from the counting rates and error rates of individual intensities.
In particular, the parameter $q^0$ can be estimated by the counting rate $p_0$
of the vacuum pulses so that we obtain $q^0=p_0-p_D$.
Once the parameter $q^0$ is fixed, 
both the AKG rates $I_{\to}$ and $I_{\leftarrow}$ are 
given as monotone increasing functions concerning $q^1 (1-\overline{h}(r^1) )$.
Hence, the remaining problem is the estimation of 
the parameters $q^1$ and $r^1$, which is a more complicated problem.

In order to calculate these parameters,
we need simultaneous equations concerning the individual intensities, 
in which 
the counting rates except for dark counts in detector and 
error rates of individual number states are treated as the unknown parameters.
For simplicity, we assume that
the counting rate
except for dark counts in detector
does not depends on the basis.
Then, we let 
$\tilde{q}^n$ be the counting rate
of the state $|n\rangle \langle n|$,
and 
$\tilde{r}^n$ be the error rate 
of the state $|n\rangle \langle n|$ with the $\times$ basis.
Note that $\tilde{q}^1=q^1$ and $\tilde{r}^1=r^1$.
Letting $p_i$ be the counting rate of the intensity $\mu_i$,
and $s_i$ be the error rate of the intensity $\mu_i$ with the $\times$ basis,
we obtain \cite{LMC}
\begin{eqnarray}
p_i & = e^{-\mu_i}(p_0-p_D)
+e^{-\mu_i}
\sum_{n=1}^\infty
\frac{\mu_i^n}{n!}\tilde{q}^n+p_D\label{4-8-1}\\
s_i p_i & = e^{-\mu_i}\frac{1}{2}(p_0-p_D)
+e^{-\mu_i}
\sum_{n=1}^\infty
\frac{\mu_i^n}{n!}\tilde{r}^n\tilde{q}^n
+\frac{1}{2}p_D
\label{4-8-2}
\end{eqnarray}
for $i=1, \ldots, k$ 
and 
\begin{eqnarray}
0 \le \tilde{q}^n \le 1 , \quad 0 \le \tilde{r}^n\le 1.\label{4-23-6}
\end{eqnarray}
That is, our task is to calculate
the 
minimum value ${I}(\vec{p},\vec{s})$ of 
$q^1 (1-\overline{h}(r^1) )$
under the conditions 
(\ref{4-8-1}), (\ref{4-8-2}), and
$1-p_D\ge \tilde{q}^n \ge 0$, $1\ge \tilde{r}^n \ge 0$,
for given vectors
$\vec{p}=(p_1,\ldots, p_k)$ and $\vec{s}=(s_1,\ldots, s_k)$.


\section{Universal upper bounds on concrete noise model}\label{s3}
In a real optical communication system,
when the intensity is $\mu_i$
the dark counts rate in the optical fiber or the detector
is $p_0$, which is equal to the counting rate of the vacuum pulse,
the counting rate $p_i$ and the error rate $s_i$ are given by
\begin{eqnarray}
&p_i =p_{i+k}=p(\mu_i):=
1- e^{-\alpha \mu_i}+ p_0 \label{10-16}\\
&s_i =s_{i+k}=s(\mu_i)
:=\frac{s(1- e^{-\alpha \mu_i})+ \frac{1}{2}p_0}{p(\mu_i) }
\label{10-16-2},
\end{eqnarray}
where 
the parameter $\alpha$ is the counting rate of the single photon 
except for dark counts in the the detector or 
interfusion on optical fiber,
and the parameter $s$ is 
the error rate of single photon for normal detection.
In an implemented channel,
the parameter is given by the function of 
the length of transmittance $L$ as
\begin{eqnarray}
\alpha  =\theta \cdot 10^{-\frac{a_1 L +a_0}{10}},
\label{4-11-1}
\end{eqnarray}
where
$\theta$ is the efficiency of the detector,
$a_0$ is loss coefficient in the detector, and
$a_1$ is loss coefficient in the communication channel.
When the detection probability
of the $n$-photon state is $1- (1-\alpha)^n+p_0$,
the relation (\ref{10-16}) holds.
When the error probability
of the $n$-photon state is 
$\frac{s(1- (1-\alpha)^n)+\frac{1}{2}p_0}{1- (1-\alpha)^n+p_0}$,
the relation (\ref{10-16-2}) holds.
Hence,
the quantities
$q^1=\alpha +p_0-p_D$ and 
$r^1=\frac{s\alpha+\frac{1}{2}(p_0-p_D)}{\alpha+p_0-p_D}$
satisfy the conditions (\ref{4-8-1}) and (\ref{4-8-2}).
Thus, we obtain
\begin{eqnarray*}
{I}(\vec{p}(\vec{\mu}),\vec{s}(\vec{\mu}))
\le 
(\alpha +(p_0-p_D))
(1-
\overline{h}(\frac{s\alpha +\frac{1}{2}(p_0-p_D)}{\alpha +(p_0-p_D)})
),
\end{eqnarray*}
where
$\vec{p}(\vec{\mu})=(p(\mu_1), \ldots, p(\mu_k))$,
$\vec{s}(\vec{\mu})=(s(\mu_1), \ldots, s(\mu_k))$.
Therefore, applying (\ref{2-16-6}) and (\ref{2-16-7}),
we obtain the following universal upper bounds of 
the AKG rates in the forward and reverse cases:
\begin{eqnarray}
\fl I_{\to}(\mu;p_0,p_D,\alpha,s)
:=&
\frac{1}{2}\Bigl[\mu e^{-\mu}
(\alpha +(p_0-p_D))
(1-\overline{h}(\frac{s\alpha +\frac{1}{2}(p_0-p_D)}{\alpha +(p_0-p_D)}))
\nonumber\\
&+ e^{-\mu}p_0
- (1-e^{-\alpha \mu}+p_0)
\eta
(\frac{s(1-e^{-\alpha \mu})+\frac{1}{2}p_0}{1-e^{-\alpha \mu}+p_0})\Bigr]
\label{10-19-10}\\
\fl I_{\leftarrow}(\mu;p_0,p_D,\alpha,s)
:=&
\frac{1}{2}\Bigl[\mu e^{-\mu}(\alpha +(p_0-p_D))
(1-\overline{h}(\frac{s\alpha +\frac{1}{2}(p_0-p_D)}{\alpha +(p_0-p_D)}))
\nonumber\\
&+ p_D- (1-e^{-\alpha \mu}+p_0)
\eta
(\frac{s(1-e^{-\alpha \mu})+\frac{1}{2}p_0}{1-e^{-\alpha \mu}+p_0})
\Bigr],
\label{10-19-11}
\end{eqnarray}
respectively.

When $p_0=p_D=0$, both rates are equal to
\begin{eqnarray*}
\frac{1}{2}(
\mu e^{-\mu}
\alpha(1-\overline{h}(s))
- (1-e^{-\alpha \mu})\overline{h}(s)),
\end{eqnarray*}
which can be approximated by 
\begin{eqnarray}
\frac{1}{2}(\mu e^{-\mu}
\alpha(1-\overline{h}(s))
- (1-e^{-\alpha \mu})\overline{h}(s))
\cong  \alpha 
\frac{\mu (e^{-\mu}- (1+e^{-\mu})\overline{h}(s))}{2}
\label{10-17}
\end{eqnarray}
when $\alpha$ is sufficiently small.

As is shown in \ref{4-13-1},
the optimum intensity is characterized as
\begin{eqnarray}
\argmax_{\mu}I_{\to}(\mu;p_0,p_D,\alpha,s)
&\le 1 \label{11-5-10}\\
\argmax_{\mu}
I_{\leftarrow}(\mu;p_0,p_D\alpha,s)
&\le 1 \label{11-5-11}.
\end{eqnarray}
If we choose intensities of decoy pulses suitably, 
the estimated parameters $q^1$ and $r^1$ are close to $\alpha$ and $s$.
Hence, 
it is suitable to choose the intensity of the 
phase-randomized coherent light producing the
raw keys among the interval $[0,1]$.

\section{Wang's expansion}\label{s4}
We review the previous results concerning the estimation of 
the parameters $q^1$ and $r^1$.
In this section, we consider only the case of $p_D=0$ because these results treated only this case.
In order to avoid infinite number of unknown parameters in (\ref{4-8-1}), 
Wang \cite{wang} proposed the following expansion of 
the states $\sum_{n=0}^\infty e^{-\mu_i}\frac{\mu^n}{n!}
|n \rangle\langle n|$ in the case of $k=2$:
\begin{eqnarray}
\fl \sum_{n=0}^\infty e^{-\mu_1}\frac{\mu_1^n}{n!}
|n \rangle\langle n|=&
e^{-\mu_1}|0\rangle\langle 0|
+e^{-\mu_1}\mu_1|1\rangle\langle 1|
+e^{-\mu_1}\mu_1^2 \Omega_2\rho_2
\label{4-9-3}
\\
\fl \sum_{n=0}^\infty e^{-\mu_2}\frac{\mu_2^n}{n!}
|n \rangle\langle n|=&
e^{-\mu_2}|0\rangle\langle 0|
+e^{-\mu_2}\mu_2|1\rangle\langle 1|
+e^{-\mu_2}\mu_2^2\Omega_2\rho_2
+e^{-\mu_2}\mu_2^2(\mu_2-\mu_1)
\Omega_3 \rho_3,
\label{4-9-4}
\end{eqnarray}
where
\begin{eqnarray}
\rho_2&:= 
\frac{1}{\Omega_2}
\sum_{n=2}^\infty 
\frac{\mu_1^{n-2}}{n!}
|n \rangle\langle n|,\\
\rho_3&:= 
\frac{1}{\Omega_3}
\sum_{n=3}^\infty \frac{\mu_2^{n-2}-\mu_1^{n-2}}{(\mu_2-\mu_1)n!}
|n \rangle\langle n|\\
\Omega_2 &:=\frac{1}{\mu_1^2}
(e^{\mu_1}-(1+\mu_1))\label{4-9-7}
\\
\Omega_3 &:=
\frac{1}{\mu_2^2}(e^{\mu_2}-(1+\mu_2+\frac{\mu_2^2}{2}))
-
\frac{1}{\mu_1^2}
(e^{\mu_1}-(1+\mu_1+\frac{\mu_1^2}{2})).\label{4-9-8}
\end{eqnarray}
Based on this expansion, 
we define the parameters $q^2,q^3,r^2$, and $r^3$ as
\begin{eqnarray}
q^j&:= \Tr \rho_j \sum_{n=2}^{\infty} \tilde{q}^n|n \rangle \langle n|
\label{4-8-10}
\\
r^j q^j&:= \Tr \rho_j \sum_{n=2}^{\infty} \tilde{q}^n\tilde{r}^n
|n \rangle \langle n|
\label{4-8-11}
\end{eqnarray}
for $j=2,3$.
Then, we have the following equations instead of (\ref{4-8-1}) and 
(\ref{4-23-6}):
\begin{eqnarray}
p_1 & = e^{-\mu_1}p_0
+e^{-\mu_1}\mu_1 q^1
+e^{-\mu_1}\mu_1^2 \Omega_2 q^2
\label{4-9-1}
\\
p_2&=
e^{-\mu_2}p_0
+e^{-\mu_2}\mu_2 q^1
+e^{-\mu_2}\mu_2^2 \Omega_2 q^2
+e^{-\mu_2}\mu_2^2 (\mu_2-\mu_1)\Omega_3 q^3
\label{4-9-1-2}\\
0 & \le q^j \le 1 \hbox{ for }j =1,2,3\label{4-23-4}.
\end{eqnarray}
Wang \cite{wang} gave a lower bound of $q^2$ in the following way.
First, 
he calculated
the maximum value $q^{2}_{U}$
of $q^2$ 
with the conditions (\ref{4-9-1}), (\ref{4-9-1-2}), and 
$q^1,q^3\ge 0$:
\begin{eqnarray*}
q^{2}_{U}=
\frac{p_1 }{\Omega_2}
(\frac{1}{\mu_2-\mu_1}
(
\frac{e^{-\mu_1} p_2}{\mu_2 e^{-\mu_2} p_1}-\frac{1}{\mu_1}
)
+\frac{e^{-\mu_1} p_0}{\mu_2 p_1})
\end{eqnarray*}
Using (\ref{4-9-1}),
he essentially obtained a lower bound $q_{2,\min}$ of $q^1$:
\begin{eqnarray*}
q_{2,\min}:=
\frac{\mu_2 e^{\mu_1}}{\mu_1(\mu_2-\mu_1)}(p_1 -e^{-\mu_1}p_0)
-\frac{\mu_1 e^{\mu_2}}{\mu_2(\mu_2-\mu_1)}(p_2-e^{-\mu_2}p_0).
\end{eqnarray*}
Wang \cite{wang2} also obtained
an upper bound $b_{1,\max}$ of $q^1 r^1$
\begin{eqnarray*}
b_{1,\max}:=
\frac{s_1 p_1 e^{\mu_1}-\frac{1}{2}p_0}{\mu_1}.
\end{eqnarray*}

On the other hand, Ma et al. \cite{XBZL} treated the case where
only two kinds of intensities are used for estimation of $q^1$ and $r^1$.
Their results can be translated into our case of $k=3$
by putting the counting rate $p_0$ of the vacuum pulse into 
the parameter $q^0$.
Assume that $\mu_1+\mu_2 < \mu_3$, $\mu_1+\mu_2<1$,
and $\mu_1<\mu_2< \mu_3$.
Then, they obtained an lower bound of $q^1$:
\begin{eqnarray}
q^1 \ge q^{1,3}_{L}
&:=\frac{\mu_3
(p_2 e^{\mu_2}-p_1 e^{\mu_1}
-
\frac{\mu_2^2-\mu_1^2}{\mu_3^2}
(p_3 e^{\mu_3}-p_0)
)
}{\mu_2\mu_3-\mu_3\mu_1
-\mu_2^2+\mu_1^2},
\end{eqnarray}
and an upper bound of $q^1 r^1$:
\begin{eqnarray}
q^1 r^1\le b^{1,2}_{U}
&:=
\frac{s_2 p_2e^{\mu_2} - s_1 p_1 e^{\mu_1}}{\mu_2-\mu_1}.
\end{eqnarray}
Taking the limit $\mu_1 \to 0$,
they obtained the 
the lower bound $q_{2,\min}$ of $q^1$ in the case of $k=2$.
Also, similarly, they independently obtain 
the upper bound $b_{1,\max}$ of $q^1 r^1$.

Consider how tight the bounds
$q_{2,\min}$ and $b_{1,\max}$ are.
As will be mentioned in Theorem \ref{th-26},
the minimum value of $q^1$ under the conditions
(\ref{4-9-1}), (\ref{4-9-1-2})
and $1\ge q^2,q^3\ge 0$
is calculated as
\begin{eqnarray*}
q^{1,2}_{\min}=\max\{q_{2,\min},q_{1,\min}\},
\end{eqnarray*}
where
\begin{eqnarray*}
q_{1,\min}:=
\frac{e^{\mu_1}}{\mu_1}(p_1- e^{-\mu_1}p_0)
- e^{-\mu_1}\mu_1^2 \Omega_2.
\end{eqnarray*}
Similar to (\ref{4-9-1}) and (\ref{4-9-1-2}), 
instead of (\ref{4-8-2}), the relations
\begin{eqnarray}
s_1 p_1 =& e^{-\mu_1}\frac{p_0}{2}
+e^{-\mu_1}\mu_1 q^1 r^1 
+e^{-\mu_1}\mu_1^2 \Omega_2 q^2 r^2
\label{4-9-2}\\
s_2 p_2=&
e^{-\mu_2}\frac{p_0}{2}
+e^{-\mu_2}\mu_2 q^1 r^1
+e^{-\mu_2}\mu_2^2 \Omega_2 q^2 r^2
+e^{-\mu_2}\mu_2^2 (\mu_2-\mu_1)\Omega_3 q^3 r^3
\label{4-9-2-2}\\
0 \le& q^j r^j \le 1 \hbox{ for }j =1,2,3\label{4-23-3}
\end{eqnarray}
hold.
Then, the maximum value 
$b^{1,2}_{\max}$
of $q^1 r^1 $ under the conditions
(\ref{4-9-2}), (\ref{4-9-2-2}),
and $1\ge q^2 r^2,q^3 r^3\ge 0$
is calculated as
\begin{eqnarray*}
b^{1,2}_{\max}
= \min\{b_{1,\max},b_{2,\max}\}
\end{eqnarray*}
where
\begin{eqnarray*}
\fl b_{2,\max}
:=
\frac{\mu_2 e^{\mu_1}}{\mu_1(\mu_2-\mu_1)}
(s_1 p_1 -\frac{1}{2}e^{-\mu_1}p_0)
-\frac{\mu_1 e^{\mu_2}}{\mu_2(\mu_2-\mu_1)}
(s_2 p_2 -\frac{1}{2}e^{-\mu_2}p_0
-e^{-\mu_2}\mu_2^2 (\mu_2-\mu_1)\Omega_3).
\end{eqnarray*}

From (\ref{4-9-2}), (\ref{4-9-2-2}),
\begin{eqnarray*}
s_2 p_2 e^{\mu_2}- s_1 p_1 e^{\mu_1}
= (\mu_2-\mu_1)q^1 r^1 +
(\mu_2^2-\mu_1^2)\Omega_2 q^2 r^2
+ \mu_2^2 (\mu_2-\mu_1)\Omega_3 q^3 r^3.
\end{eqnarray*}
Since $(\mu_2^2-\mu_1^2)\Omega_2 q^2 r^2\ge 0$ and 
$\mu_2^2 (\mu_2-\mu_1)\Omega_3 q^3 r^3\ge 0$,
we obtain $b_U^{1,3} \ge q^1 r^1$. Hence, 
$b_U^{1,3} \ge b^{1,2}_{\max}$, i.e., 
the bound $b^{1,2}_{\max}$ is better than $b_U^{1,3}$.

Wang \cite{wang2} proposed that
the vacuum pulse and two kinds of intensities are 
used for estimating the parameters,
and another intensity is used for signal pulse.
In this method, it is possible to use the counting rate and 
the error rate of the signal pulse.
Hence, it can be expected to improve the AKG rate
by taking into account the the counting rate and 
the error rate of the signal pulse.
That is, it is needed to discuss the case of $k=3$.
In this case, Ma et al.'s bound $q^{1,3}_{L}$ can be applied.
There is a possibility to improve existing bounds 
by extending Wang's expansion (\ref{4-9-3}) and (\ref{4-9-4}) 
to the case of $k=3$.
Further, as was pointed out by Ma et al.\cite{XBZL},
we can expect that 
the AKG rates are close to the upper bounds 
$I_{\to}(\mu;p_0,p_D,\alpha,s)$ and 
$I_{\leftarrow}(\mu;p_0,p_D,\alpha,s)$
if the number $k$ of intensities is sufficiently large.
In the following section, we concentrate to derive better estimates
of $q^1$ and $r^1 q^1$.

\section{Convex expansion of mixed state $\sum_{n=0}^{\infty}\frac{e^{-\mu}\mu^n}{n!}|n\rangle \langle n|$}\label{s5}
\subsection{Extension of Wang's expansion}
In this section, 
we give a convex expansion of 
the phase-randomized coherent state
$\sum_{n=0}^{\infty}\frac{e^{-\mu}\mu^n}{n!}
|n\rangle \langle n|$ as an extension of 
Wang's expansion (\ref{4-9-3}) and (\ref{4-9-4}):
\begin{thm}\label{25-1}
Assume that $\mu_1 < \ldots < \mu_k$.
Define the basis state $\rho_i$ ($i=2,\ldots, k+1$) as
\begin{eqnarray*}
\rho_i&:=
\frac{1}{\Omega_i}
\sum_{n=i}^\infty
\frac{\gamma_{i,n}}{n!}
|n \rangle \langle n|,\\
\gamma_{i,n}&:=
\sum_{j=1}^{i-1}
\frac
{\mu_j^{n-2}}
{\prod_{t=1,t\neq j}^{i-1}(\mu_j-\mu_{t})}&
\\
\Omega_i&:=\sum_{n=i}^\infty\frac{\gamma_{i,n}}{n!}.
\end{eqnarray*}
Then, 
$\rho_i$ is positive semi-definite, and
\begin{eqnarray}
e^{-\mu_i}\sum_{n=0}^\infty \frac{\mu_i^n}{n !}|n\rangle \langle n|
=
e^{-\mu_i}
\Bigl(|0\rangle \langle 0|
+\mu_i |1\rangle \langle 1|
+
\sum_{n=2}^{i+1}
\mu_i^2 \prod_{t=1}^{n-2}(\mu_i-\mu_t)
\Omega_n 
\rho_n
\Bigr)\label{10-20-2}.
\end{eqnarray}
Note that the coefficient $
\mu_i^2 \prod_{t=1}^{n-2}(\mu_i-\mu_t)
\Omega_n $ is positive.
\end{thm}
The quantities $\Omega_2$ and $\Omega_3$
coincide with those defined in (\ref{4-9-7}) and (\ref{4-9-8}).
Hence, we can check that 
the above expansion of the case of $k=2$ 
reproduces Wang's expansion (\ref{4-9-3}) and (\ref{4-9-4}).
Further, $\Omega_4$ and $\Omega_5$
are calculated as
\begin{eqnarray*}
\Omega_4 
=&
\frac{1}{\mu_3^2(\mu_3-\mu_1)}
(e^{\mu_3}-1 - \mu_3 - \frac{\mu_3^2}{2}- \frac{\mu_3^3}{6})
-
\frac{1}{\mu_2^2(\mu_2-\mu_1)}
(e^{\mu_2}-1 - \mu_2 - \frac{\mu_2^2}{2}- \frac{\mu_2^3}{6})\\
&+
\frac{1}{\mu_1^2(\mu_2-\mu_1)}
(e^{\mu_1}-1 - \mu_1 - \frac{\mu_1^2}{2}- \frac{\mu_1^3}{6})
\end{eqnarray*}
and
\begin{eqnarray*}
&\Omega_5 \\
=&
\frac{1}{\mu_4^2(\mu_4-\mu_2)(\mu_4-\mu_1)}
(e^{\mu_4}-1 - \mu_4 - \frac{\mu_4^2}{2}- \frac{\mu_4^3}{6}
- \frac{\mu_4^4}{24})\\
& -
\frac{1}{\mu_3^2(\mu_3-\mu_2)(\mu_3-\mu_1)}
(e^{\mu_3}-1 - \mu_3 - \frac{\mu_3^2}{2}- \frac{\mu_3^3}{6}
- \frac{\mu_3^4}{24})
\nonumber
\\
& +
\frac{1}{\mu_2^2(\mu_3-\mu_2)(\mu_2-\mu_1)}
(e^{\mu_2}-1 - \mu_2 - \frac{\mu_2^2}{2}- \frac{\mu_2^3}{6}
- \frac{\mu_2^4}{24})\nonumber
\\
& -
\frac{1}{\mu_1^2(\mu_3-\mu_1)(\mu_2-\mu_1)}
(e^{\mu_1}-1 - \mu_1 - \frac{\mu_1^2}{2}- \frac{\mu_1^3}{6}
- \frac{\mu_1^4}{24}).
\end{eqnarray*}

In order to estimate the parameters $q^1$ and $r^1$,
we introduce new parameters 
$q^2, \ldots, q^{k+1}$, and 
$r^2, \ldots, r^{k+1}$ by
\begin{eqnarray}
q^j&:= \Tr \rho_j \sum_{n=2}^{\infty} \tilde{q}^n|n \rangle \langle n|
\label{4-9-20}
\\
r^j q^j&:= \Tr \rho_j \sum_{n=2}^{\infty} \tilde{q}^n\tilde{r}^n
|n \rangle \langle n|
\label{4-9-21}
\end{eqnarray}
for $j=2,\ldots, k+1$.
Hence, instead of (\ref{4-8-1}) and (\ref{4-8-2}), as a generalization of 
(\ref{4-9-1}), (\ref{4-9-1-2}), (\ref{4-9-2}), and (\ref{4-9-2-2}) 
we obtain the equations:
\begin{eqnarray*}
p_i &= \sum_{j=0}^{k+1} {P'}_i^j q^j, 
\\
s_i p_i &= \sum_{j=1}^{k+1} {P'}_i^j q^j r^j+\frac{1}{2} {P'}_i^0 q^0,
\end{eqnarray*}
for $i=1,2,\ldots,k$,
where
\begin{eqnarray*}
P':=\left(
\begin{array}{ccc}
1 &0& 0 \\
Y &Z& X 
\end{array}
\right),
\end{eqnarray*}
and the $k$-dimensional vectors $Y$ and $Z$ and 
the $k \times k$ matrix $X$ are given by
\begin{eqnarray*}
Y_i &:=e^{-\mu_i}\\
Z_i &:=\mu_i e^{-\mu_i} \\
X_i^j&:=
\left\{
\begin{array}{ll}
\mu_i^2
\prod_{t=1}^{j-1} (\mu_i-\mu_t)
e^{-\mu_i}\Omega_{j+1}
& \hbox{if }j= 1,\ldots, i \\
0 & \hbox{if }j = i+1,\ldots, k
\end{array}
\right.
\end{eqnarray*}
for $i=1,\ldots, k$.

\subsection{General case}
In the previous subsection, we treat the case where
$p_D=0$ and the counting rates and the error rates
of the $\times$ basis are equal to those of the $+$ basis.
Since the average state of the state $\rho_j$ with the $\times$ basis 
is different from that the $+$ basis,
we have to treat the parameter 
$q^{j+k}$ of the $+$ basis
as a parameter different from the parameter $q^j$ of the $\times$ basis
in the general setting.
In this subsection, we extend the discussion of the above subsection
to the general case.
Let $\tilde{q}^n_{\times}$ and $\tilde{r}^n_{\times}$
be the counting rate except for dark counts in detector
and error rate of the number state
$|n \rangle \langle n|$ with the $\times$ basis,
and 
$\tilde{q}^n_{+}$ and $\tilde{r}^n_{+}$
be those with the $+$ basis.
Instead of (\ref{4-9-20}) and (\ref{4-9-21}),
the parameters 
$q^2, \ldots, q^{2k+1}$, and
$r^2, \ldots, r^{k+1}$ 
are introduced by
\begin{eqnarray}
q^j&:= \Tr \rho_j \sum_{n=2}^{\infty} \tilde{q}^n_{\times}|n \rangle \langle n|
\label{4-9-22}
\\
q^{j+k}&:= \Tr \rho_j \sum_{n=2}^{\infty} \tilde{q}^n_{+}|n \rangle \langle n|
\label{4-9-23}\\
r^j q^j&:= \Tr \rho_j \sum_{n=2}^{\infty} \tilde{q}^n_{\times}
\tilde{r}^n_{\times}|n \rangle \langle n|
\label{4-9-24}
\end{eqnarray}
for $j=2,\ldots, k+1$.
That is, 
the upper index $j$ has the following meaning:
\begin{description}
\item[$j=0$:] The vacuum state 
\item[$j=1$:] The single-photon state
\item[$j=2, \ldots, k+1$:] The state 
$\rho_j$ with the $\times$ basis 
\item[$j=k+2, \ldots, 2k+1$:]
The state $\rho_{j-k}$ with the $\times$ basis 
\end{description}
Note that
$q^j$ is the rate of counting except for dark counts.
Instead of (\ref{4-9-22}) -- (\ref{4-9-24}),
we have the relations
\begin{eqnarray}
p_i &= \sum_{j=0}^{2k+1} P_i^j q^j+p_D, 
\label{9-15-2-2}
\end{eqnarray}
for $i=0, \ldots, {2k}$, and
\begin{eqnarray}
s_i p_i &= \sum_{j=1}^{k+1} P_i^j q^j r^j+\frac{1}{2} (P_i^0 q^0 + p_D)
\label{9-15-1-2},
\end{eqnarray}
for $i=1,2,\ldots,k$.
In addition, $r^i$ belongs to $[0,1]$ for $i=1,\ldots, k$,
and $q^i$ belongs to $[0,1-p_D]$ for $i=0,\ldots, 2k+1$
because the dark counts occur with the probability $p_D$.
Therefore, we can estimate of the ranges of $q^1$ and $r^1$ 
from these conditions.
Here, we defined the matrix 
$(P_{i}^j)_{i=0, \ldots, 2k, j= 0, \ldots, 2k+1}$ 
defined by 
\begin{eqnarray*}
P:=\left(
\begin{array}{cccc}
1 &0& 0 & 0 \\
Y &Z& X & 0 \\
Y &Z& 0 & X
\end{array}
\right),
\end{eqnarray*}
where the $k$-dimensional vectors $Y$ and $Z$ and 
the $k \times k$ matrix $X$ are given the above.

For examples, this matrix in the case of $k=1,2,3$ 
is given as 
\begin{eqnarray*}
\fl P_1&=
\left(
\begin{array}{cccc}
1 & 0 & 0 & 0 \\
e^{-\mu_1} & \mu_1 e^{-\mu_1} & e^{-\mu_1}\mu_1^2 \Omega_2 & 0 \\
e^{-\mu_1} & \mu_1 e^{-\mu_1} & 0 & e^{-\mu_1}\mu_1^2 \Omega_2
\end{array}
\right),\\
\fl P_2&=
\left(
\begin{array}{cccccc}
1 & 0 & 0 & 0 & 0 & 0 \\
e^{-\mu_1} & \mu_1 e^{-\mu_1} & e^{-\mu_1}\mu_1^2 \Omega_2& 0 & 0 & 0\\
e^{-\mu_2} & \mu_2 e^{-\mu_2} & e^{-\mu_2}\mu_2^2 \Omega_2
& e^{-\mu_2}\mu_2^2(\mu_2-\mu_1)\Omega_3 & 0 & 0\\
e^{-\mu_1} & \mu_1 e^{-\mu_1} & 0 & 0 & e^{-\mu_1}\mu_1^2 \Omega_2 & 0 \\
e^{-\mu_2} & \mu_2 e^{-\mu_2} & 0 & 0 &
e^{-\mu_1}\mu_2^2 \Omega_2& e^{-\mu_2}\mu_2^2(\mu_2-\mu_1)\Omega_3 
\end{array}
\right),\\
\fl P_3&=\left(
\begin{array}{cccccccc}
1 & 0 & 0 & 0 & 
0 & 0 & 0 & 0 \\
e^{-\mu_1} & \mu_1 e^{-\mu_1} & e^{-\mu_1}\mu_1^2\Omega_2 & 0 &
0 & 0 & 0 & 0\\
e^{-\mu_2} & \mu_2 e^{-\mu_2} & e^{-\mu_2}\mu_2^2\Omega_2
& e^{-\mu_2}\mu_2^2(\mu_2-\mu_1)\Omega_3 & 0 & 0 & 0 & 0\\
e^{-\mu_3} & \mu_3 e^{-\mu_3} & e^{-\mu_3} \mu_3^2 \Omega_2 
& e^{-\mu_3}\mu_3^2(\mu_3-\mu_1)\Omega_3 
& *
& 0 & 0 & 0 \\
e^{-\mu_1} & \mu_1 e^{-\mu_1} & 0 & 0 & 0 & e^{-\mu_1}\mu_1^2\Omega_2 & 0 & 0\\
e^{-\mu_2} & \mu_2 e^{-\mu_2} & 0 & 0 & 0 &
e^{-\mu_2} \mu_2^2\Omega_2 &  e^{-\mu_2}\mu_2^2(\mu_2-\mu_1)\Omega_3
& 0 \\
e^{-\mu_3} & \mu_3 e^{-\mu_3} & 0 & 0 & 0 &
e^{-\mu_3}\mu_3^2 \Omega_2& 
e^{-\mu_3} \mu_3^2(\mu_3-\mu_1)\Omega_3 
& *
\end{array}
\right)
\end{eqnarray*}
where
\begin{eqnarray*}
*&=e^{-\mu_3}(\mu_3-\mu_1)(\mu_3-\mu_2)\Omega_4 .
\end{eqnarray*}

\section{Asymptotic key generation (AKG) rates}\label{s6}
In this section, 
based on the expansion (\ref{10-20-2}),
we derive a
lower bound of 
$I(\vec{p},\vec{s})$.
Since the matrix $P_i^j$ has no inverse matrix,
it is impossible to derive $q^1$ and $r^1$ from the conditions 
(\ref{9-15-2-2}) and (\ref{9-15-1-2}) uniquely.
Then, 
in order to evaluate $I(\vec{p},\vec{s})$,
we introduce 
the quantities 
$q^{1,k}_{\min}$ and $b^{1,k}_{\max}$ by
\begin{eqnarray}
q^{1,k}_{\min}
:=
& \min_{\vec{q}=(q^0,\ldots,q^{2k+1})^T}
\left\{q^1\left|
\begin{array}{l}
p_i = \sum_{j=0}^{2k+1} P_i^j q^j+p_D\\
1-p_D \ge q^0,\ldots,q^{2k+1} \ge 0
\end{array}
\right.\right\}\nonumber \\
b^{1,k}_{\max}
:=&
\max_{\vec{b}=(b^1,\ldots,b^{k+1})^T}
\left\{b^1\left|
\begin{array}{l}
s_i p_i = \sum_{j=1}^{k+1} P_i^j b^j\\
\hspace{7ex}+\frac{P_i^0 (p_0-p_D) + p_D}{2} \\
1-p_D\ge b^1, \ldots ,b^{k+1}\ge 0 
\end{array}
\right.\right\}\nonumber 
\end{eqnarray}
for $i= 0,\ldots, k$.
Then, 
the minimum value $I(\vec{p},\vec{s})$ can be evaluated by
\begin{eqnarray*}
 q^{1,k}_{\min}(1- \overline{h}(
\frac{b^1_{\max}}{q^{1,k}_{\min}}
))
\le 
I(\vec{p},\vec{s}) .
\end{eqnarray*}
Thus, when $\mu_k$ is the signal intensity,
\begin{eqnarray*}
\frac{1}{2}
\left(
\mu_k e^{-\mu_k}
I(\vec{p},\vec{s}) 
+e^{-\mu_k}(q^0+p_D)
- p_{\mu_k}\eta(s_{\mu_k})
\right)
&\le I_{\to}\\
\frac{1}{2}\left(
\mu_k e^{-\mu_k} I(\vec{p},\vec{s}) 
+p_D- p_{\mu_k}\eta(s_{\mu_k})
\right)
& \le
I_{\leftarrow}.
\end{eqnarray*}

In the following, we calculate 
$b^{1,k}_{\max}$ and $q^{1,k}_{\min}$ as follows.
\begin{thm}\label{th-26}
Define the quantities $q_{j,\min}$, $q_{k+j,\min}$, and $b_{j,\max}$ by
\begin{eqnarray}
\fl q_{j,\min}:=
\left\{
\begin{array}{ll}
\sum_{i=1}^j \beta^j_i (p_i-p_D-e^{-\mu_i}(p_0-p_D))
-(1-p_D)\mu_1 \cdots \mu_j\Omega_{j+1}
& \hbox{if }j \hbox{ is odd.}\\
\sum_{i=1}^j \beta^j_i (p_i-p_D-e^{-\mu_i}(p_0-p_D))
& \hbox{if }j \hbox{ is even.}
\end{array}
\right.
\label{11-12-5}\\
\fl q_{k+j,\min}:=
\left\{
\begin{array}{ll}
\sum_{i=1}^j \beta^j_i (p_{i+k}-p_D-e^{-\mu_i}(p_0-p_D))
-(1-p_D)\mu_1 \cdots \mu_j\Omega_{j+1}
& \hbox{if }j \hbox{ is odd.}\\
\sum_{i=1}^j \beta^j_i (p_{i+k}-p_D-e^{-\mu_i}(p_0-p_D))
& \hbox{if }j \hbox{ is even.}
\end{array}
\right.
\label{11-12-5-1}\\
\fl b_{j,\max}:=
\left\{
\begin{array}{ll}
\sum_{i=1}^j \beta^j_i (s_i p_i-\frac{1}{2}(p_D+e^{-\mu_i}(p_0-p_D)))
& \hbox{if }j \hbox{ is odd.}\\
\sum_{i=1}^j \beta^j_i (s_i p_i-\frac{1}{2}(p_D+e^{-\mu_i}(p_0-p_D)))
+(1-p_D)\mu_1 \cdots \mu_j\Omega_{j+1}
& \hbox{if }j \hbox{ is even.}
\end{array}
\right.
\label{11-12-8}\\
\fl \beta^j_i
:=
(-1)^{j-1}
\frac{\mu_1\cdots \mu_j e^{\mu_i}}
{\mu_i^2 
\prod_{t=1,t\neq i}^{j}(\mu_i-\mu_t)}
\label{4-27-2}
\end{eqnarray}
for $j \le k$.
Then, the relations
\begin{eqnarray}
q_{j,\min}=
& \min_{\vec{q}=(q^0,\ldots,q^{2k+1})^T}
\left\{q^1\left|
\begin{array}{l}
p_i = \sum_{j=0}^{2k+1} P_i^j q^j+p_D\\
1-p_D\ge q^{1+j}\ge 0
\end{array}
\right.\right\}\label{11-12-1} \\
q_{k+j,\min}=
& \min_{\vec{q}=(q^0,\ldots,q^{2k+1})^T}
\left\{q^1\left|
\begin{array}{l}
p_i = \sum_{j=0}^{2k+1} P_i^j q^j+p_D\\
1-p_D\ge q^{1+j+k} \ge 0
\end{array}
\right.\right\}\label{11-12-2} \\
b_{j,\max}
=&
\max_{\vec{b}=(b^1,\ldots,b^{k+1})^T}
\left\{b^1\left|
\begin{array}{l}
s_i p_i = \sum_{j=1}^{k+1} P_i^j b^j+\frac{P_i^0 (p_0-p_D) + p_D}{2} \\
1-p_D\ge b^{1+j}\ge 0
\end{array}
\right.\right\}\label{11-12-4},
\end{eqnarray}
hold.
Therefore, $q^{1,k}_{\min}$ and $b^{1,k}_{\max}$ are calculated as
\begin{eqnarray}
q^{1,k}_{\min}&= \max\{q_{1,\min},\ldots,q_{2k,\min}\}\label{10-26-1}
\\
b^{1,k}_{\max}&= \min\{b_{1,\max},\ldots,b_{k,\max}\}.\label{10-26-4}
\end{eqnarray}
\end{thm}

In order to calculate $q_{j,\min}$ and $b_{j,\max}$ 
in the channel model (\ref{10-16}) and (\ref{10-16-2}),
we define the quantity $\epsilon^j_\alpha(\mu_1, \ldots, \mu_j)$:
\begin{eqnarray}
\epsilon^j_\alpha(\mu_1, \ldots, \mu_j):=
(-1)^{j-1}(\sum_{i=1}^j \beta_i^j (1-e^{-\alpha \mu_i})-\alpha)
\ge 0.\label{4-18-1}
\end{eqnarray}
This quantity can be characterized by the following theorem.
\begin{thm}\label{th-26-2-a}
We denote the $j-1$-dimensional simplex 
and its uniform probability measure
by $\Delta_{j-1}$ and $p_{j-1}$, respectively.
Then, $\epsilon^j_\alpha(\mu_1, \ldots, \mu_j)$ is characterized as follows.
\begin{eqnarray}
\fl \epsilon^j_\alpha(\mu_1, \ldots, \mu_j)\nonumber\\
\fl =
\frac{\mu_1\cdots\mu_j}{(j-1)!}
\int_{\Delta_{j-1}}
\sum_{n=0}^\infty
\frac{1-(1-\alpha)^{n+j+1}}{(n+j+1)(n+j)n!}
(\sum_{i=1}^{j}a_i \mu_i)^{n}
p_{j-1}(a_1\ldots a_{j})d a_1\ldots d a_{j}\label{11-4-1}\\
\fl =
\mu_1 \cdots \mu_j
\sum_{n=j+1}^{\infty}
\sum_{i_l\ge0:i_1+\cdots+i_j=n-1-j}
\frac{1}{n!}(1-(1-\alpha)^n)
\mu_1^{i_1}\cdots\mu_1^{i_j}.\label{4-23-2}
\end{eqnarray}
$\Omega_{j+1}$ is calculated as
\begin{eqnarray}
\mu_1 \cdots \mu_j\Omega_{j+1}=
\epsilon_{1}^j(\mu_1,\ldots,\mu_j ).
\label{10-26-11}
\end{eqnarray}
\end{thm}

Using the relations (\ref{4-18-1}) and (\ref{10-26-11}),
we can calculate $q_{j,\min}$, $q_{k+j,\min}$,
 and $b_{j,\max}$ in the 
channel model
(\ref{10-16}) and (\ref{10-16-2}) as follows:
\begin{eqnarray*}
\fl q_{j,\min} =q_{k+j,\min} 
\left\{
\begin{array}{ll}
\alpha +(p_0-p_D)
+ \epsilon^j_\alpha(\mu_1, \ldots, \mu_j)
-(1-p_0)\epsilon^j_1(\mu_1, \ldots, \mu_j)
&\hbox{if } j \hbox{ is odd.}\\
\alpha +(p_0-p_D)
- \epsilon^j_\alpha(\mu_1, \ldots, \mu_j)
-(p_0-p_D)\epsilon^j_1(\mu_1, \ldots, \mu_j)
&\hbox{if } j \hbox{ is even.}
\end{array}
\right.\\
\fl b_{j,\max}=
\left\{
\begin{array}{ll}
s\alpha +\frac{1}{2}(p_0-p_D)
+ s\epsilon^j_\alpha(\mu_1, \ldots, \mu_j)
+\frac{p_0-p_D}{2}\epsilon^j_1(\mu_1, \ldots, \mu_j)
&\hbox{if } j \hbox{ is odd.}\\
s\alpha +\frac{1}{2}(p_0-p_D)
- s\epsilon^j_\alpha(\mu_1, \ldots, \mu_j)
+(1-\frac{1}{2}(p_0+p_D))\epsilon^j_1(\mu_1, \ldots, \mu_j)
&\hbox{if } j \hbox{ is even.}
\end{array}
\right.
\end{eqnarray*}
From the expression (\ref{11-4-1}),
$\epsilon^j_\alpha(\mu_1, \ldots, \mu_j)$
is monotone increasing concerning 
all of $\mu_1, \ldots, \mu_j$, and $\alpha$.
Also the expression (\ref{11-4-1}) implies that
$\epsilon^j_\alpha(\mu_1, \ldots, \mu_j)$ goes to $0$ when $\mu_1$ goes to $0$.
That is, the upper bounds 
$I_{\to}(\mu;p_0,p_D,\alpha,s)$ and $I_{\leftarrow}(\mu;p_0,p_D,\alpha,s)$
can be attained.
This fact coincides with the fact that
$\tilde{q}^{1,2}_L$ goes to $s$ when $\mu_1$ goes to $0$,
which was proved by Ma et al.\cite{XBZL}.
However, in a realistic system,
it is impossible to take the limit $\mu_1\to 0$.
Even if, we could realize such a small $\mu_1$,
the estimation process is not robust for statistical fluctuation.
Hence, 
in order to attain the upper bounds 
$I_{\to}(\mu;p_0,p_D,\alpha,s)$ and $I_{\leftarrow}(\mu;p_0,p_D,\alpha,s)$,
it is suitable to fix the minimum of the width $\mu_i-\mu_{i-1}$ and 
increase the number $k$.

\section{Comparison of AKG rates}\label{s7}
In order to compare estimates of 
$q^1$ and $q^1 r^1$,
we assume that $p_i=p_{i+k}$.
As is mentioned in Section \ref{s4},
$q^{1,2}_{\min}$ and $b^{1,2}_{\max}$ 
give the best estimates of 
$q^1$ and $q^1 r^1$ among known estimates
in the case of $k=2$.

In this section, 
we assume the channel model (\ref{10-16})-(\ref{4-11-1}).
As a typical case,
we focus on the case of 
$a_1=$0.17 dB/km,
which is  
the lowest loss values in commercially available optical fibres\cite{Loss}
$a_0=$5dB,
$\theta=0.1$,
$p_0=4.0\times 10^{-7}$,
$s=0.03$\cite{KNHTKN04}.

The minimum of the width
$\mu_i-\mu_{i-1}$ is $0.1$.
For simplicity, 
we assume that our code of classical error correction attains 
the Shannon rate.

\begin{figure}[htbp]
\begin{center}
\includegraphics[width=8cm]{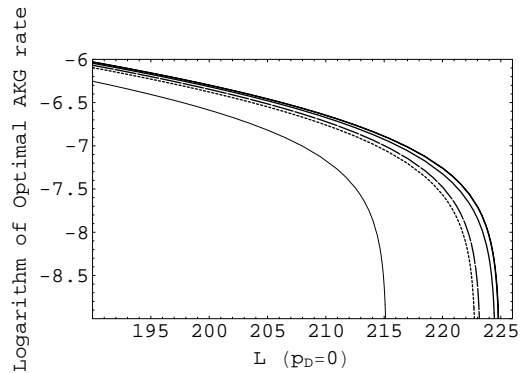}
\end{center}
\caption{The transmission rate ($p_D=0$):
From left to right,
(3.1), (2), (3.2), (3.3), (4), (5).
We cannot dintingush (4) and (5) in this graph.
}
\label{PD=0}
\end{figure}

Now, we compare the AKG rates with the forward error correction
with exsiting estimates.
For fair comparsion, we do not take into account the dark count effect,
i.e., assume that $p_D=0$.
Then, we calculate the following values as functions of the distance $L$:
\begin{description}
\item[(2)]
$\max_{0.2<\mu_2}I(\mu_2,q_{2,\min}(0.1,\mu_2),
b_{1,\max}(0.1))$.Ma et al.\cite{XBZL}, Wang\cite{wang2}
\item[(3.1)]
$\max_{0.3<\mu_3}I(\mu_3,q^{1,3}_{L}(0.1,0.2,\mu_3),
b_{1,\max}(0.1))$.Ma et al.\cite{XBZL}
\item[(3.2)]$\max_{0.3<\mu_3}I(\mu_3,q_{2,\min}(0.1,0.2),
b_{1,\max}(0.1))$.Wang\cite{wang2} $k=3$
\item[(3.3)]
$\max_{0.3<\mu_3}I(\mu_3,q^{1,3}_{\min}(0.1,0.2,\mu_3),
b^{1,3}_{\max}(0.1,0.2,\mu_3))$. Our result $k=3$
\item[(4)]$\max_{0.4<\mu_4}I(\mu_4,q^{1,4}_{\min}(0.1,0.2,0.3,\mu_4),
b^{1,4}_{\max}(0.1,0.2,0.3,\mu_4))$. Our result $k=4$
\item[(5)]$\max_{0<\mu}I(\mu,\alpha-p_0+p_D,s \alpha +\frac{1}{2}(p_0-p_D))$. 
Upper bound
\end{description}
Here, in order to treat the forward error correction case with $p_D=0$,
we put $I(\mu,q^1,b^1)$ as
$I(\mu,q^1,b^1):=
\frac{1}{2}(\mu e^{-\mu} q^1 (1-h(\frac{b^1}{q^1}))+e^{-\mu}p_0 -p(\mu)h(s(\mu)))$.
In this case,
we can numerically check that
$q^{1,2}_{\min}=q_{2,\min}$,
$q^{1,3}_{\min}=q_{2,\min}$,
$q^{1,4}_{\min}=q_{4,\min}$,
$b^{1,2}_{\max}=b_{1,\max}< b^{1,2}_{U}$,
$b^{1,3}_{\max}=b_{3,\max}$,
$b^{1,4}_{\max}=b_{3,\max}$.
Here, we treat $q^{1,3}_{L}$, $q^{1,2}_{L}$,
$q_{2,\min}$, $q^{1,3}_{\min}$,
$b_{1,\max}$, $b^{1,2}_{\max}$, $b^{1,3}_{\max}$, 
as functions of $\mu_1,\mu_2$, ($\mu_3$, $\mu_4$) 
with the model (\ref{10-16}), (\ref{10-16-2}).
The transmission rates of the abobve six cases
are given in the Fig \ref{PD=0}.
The acheivable transmission length is
(2)222.8km,
(3.1)215.2km,
(3.2)223.2km,
(3.3)224.5km,
(4)224.8km,
(5)225.2km.
That is, by incereasing the number $k$ from $2$ to $3$
yields increasing the acheivable transmission length increases with 1.7 km,
while incereasing the number $k$ from $3$ to the infinity
yields increasing it only with 0.7 km.

Next, taking into account the dark count effect,
we consider the effect of increase of the number $k$ of intensities
with the forward and reverse error correction.
In these comparsions, it is assumed that
all detections with the vacuum pulse 
are caused by the dark count effect in the detector, i.e., $p_D=p_0$.

\begin{figure}[htbp]
\begin{center}
\includegraphics[width=8cm]{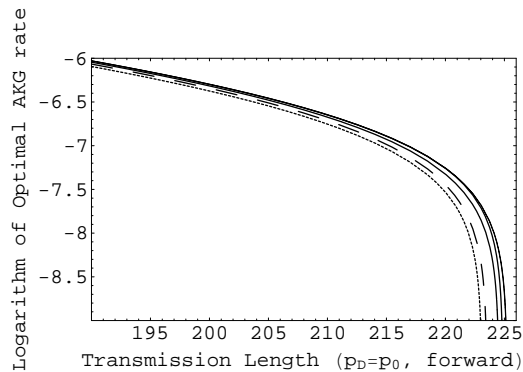}
\end{center}
\caption{The transmission rate
with the forward error correction 
($p_D=p_0$):
From left to right,
(2), (3.2), (3.3), (4), (5)}
\label{forward}
\end{figure}

In order to discuss the forward error correction case with $p_D=p_0$,
we replace 
the definitions of 
$q_{2,\min}$,
$b_{1,\min}$,
$q^{1,3}_{\min}$,
$b^{1,3}_{\min}$,
$q^{1,4}_{\min}$, and
$b^{1,4}_{\min}$
in the above table
because these depend on the value $p_D$.
The transmission rates of the five cases
(2), (3.2), (3.3), (4), and (5) with the forward error correction
are given in the Fig \ref{forward}.
The acheivable transmission length 
in the forward case is
(2)223.0km,
(3.2)223.5km,
(3.3)224.5km,
(4)224.8km,
(5)225.2km.
That is, incereasing the number $k$ from $2$ to $3$
yields increasing the acheivable transmission length increases with 1.5 km,
while incereasing the number $k$ from $3$ to the infinity
yields increasing it only with 0.7 km.

\begin{figure}[htbp]
\begin{center}
\includegraphics[width=8cm]{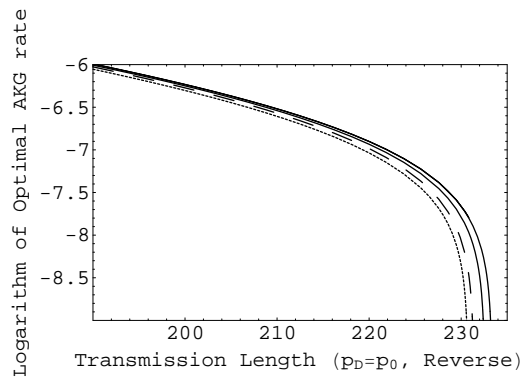}
\end{center}
\caption{The transmission rate with the reverse error correction 
($p_D=p_0$):
From left to right,
(2), (3.2), (3.3), (4), (5).
We cannot dintingush (4) and (5) in this graph.}
\label{reverse}
\end{figure}

In order to discuss the reverse error correction case with $p_D=p_0$,
we replace the definition of $I(\mu,q^1,b^1)$ as
$I(\mu,q^1,b^1):=
\frac{1}{2}(\mu e^{-\mu} q^1 (1-h(\frac{b^1}{q^1}))+p_D -p(\mu)h(s(\mu)))$
in the above table.
The transmission rates of the five cases
(2), (3.2), (3.3), (4), and (5) with the reverse error correction
are given in the Fig \ref{reverse}.
The acheivable transmission length 
in the forward case is
(2)230.7km,
(3.2)231.3km,
(3.3)232.5km,
(4)233.2km,
(5)233.3km.
That is, incereasing the number $k$ from $2$ to $3$,
yields increasing the acheivable transmission length increases with 1.8 km,
while incereasing the number $k$ from $3$ to the infinity
yields increasing it only with 0.8 km.
These comparisions indicate that
the AKG rate is almost saturated in the case of $k=3$.
Further, these graphs (Figs \ref{PD=0}, \ref{forward}, and \ref{reverse})
show that
our AKG rate $\max_{0.3<\mu_3}I(\mu_3,q^{1,3}_{\min}(0.1,0.2,\mu_3),
b^{1,3}_{\max}(0.1,0.2,\mu_3))$
is better than Wang's proposal in the case of $k=3$.

\begin{figure}[htbp]
\begin{center}
\includegraphics[width=8cm]{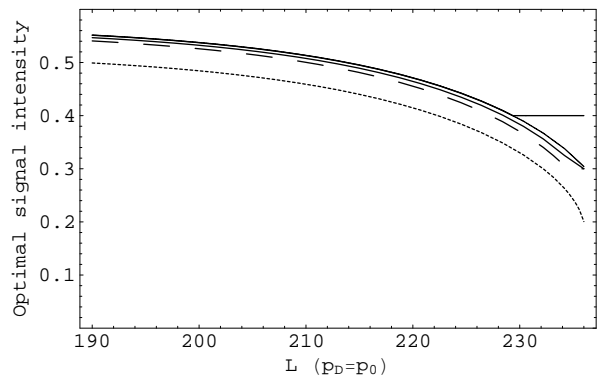}
\end{center}
\caption{The optimal signal intensity with the reverse error correction 
($p_D=p_0$):
From up to down,
(4), (5), (3.3), (3.2), (2)}
\label{intensity}
\end{figure}

The optimal signal intensity with the reverse error correction 
is calculated as Fig \ref{intensity}.

\section{Proofs of Theorems}\label{s8}
\subsection{Generalization of the concept ``difference''}
In this section, 
in order to prove theorems given in above sections,
we generalize the concept of ``difference''.
In the conventional mean value theorem,
we focus on the difference $\frac{f(x_2)-f(x_1)}{x_2-x_1}$
between two points $x_1$ and $x_2$ for a given real-valued function $f$.
When we treat $n+1$ points $x_1, \ldots, x_{n+1}$,
we introduce the quantity $
\Delta^{n+1}_f(x_1, \ldots, x_{n+1}):=
\sum_{i=1}^{n+1} 
\frac{f(x_i)}{\prod_{1\le j \le n+1, j \neq i}(x_i-x_j)}$ 
as a generalization of the difference.
This generalized difference is characterized by 
$n$-th derivative by
the following generalization of the mean value theorem:
\begin{lem}\label{th-23}
Any $n$-differentiable function $f$ on $\real$
and any $n+1$ points $x_1< \ldots < x_{n+1}$
satisfy the equation:
\begin{eqnarray}
\fl
\Delta^{n+1}_f(x_1, \ldots, x_{n+1})
=
\frac{1}{n!}
\int_{\Delta_{n}}
f^{(n)}(\sum_{i=1}^{n+1}a_i x_i)
p_n(a_1\ldots a_{n+1})d a_1\ldots d a_{n+1},
\label{10-23-1}
\end{eqnarray}
where $\Delta_{n}$ is the $n$-th simplex,
and $p_n$ is the standard uniform probability measure on $\Delta_n$.
Hence, 
there exists a point $z \in [x_1,x_{n+1}]$ such that
\begin{eqnarray*}
f^{(n)}(z)=
n!
\Delta^{n+1}_f(x_1, \ldots, x_{n+1}).
\end{eqnarray*}
\end{lem}
Its proof will be given in \ref{a1}.
Therefore, the quantity 
$\Delta^{n+1}_f(x_1, \ldots, x_{n+1})$
can be regarded as a generalization of difference.
This quantity satisfies the following recurrence formula.
\begin{lem}\label{lem1}
\begin{eqnarray}
\Delta^{n+1}_f(x_1, \ldots, x_{n+1})
=
\frac{\Delta^{n}_f(x_2, \ldots, x_{n+1})
-\Delta^{n}_f(x_1, \ldots, x_{n})
}{x_{n+1}-x_1}
.\label{4-20-1}
\end{eqnarray}
\end{lem}
Its proof will be given in \ref{a2}.

Using this formula, we can prove
the following by induction:
\begin{eqnarray}
\fl \Delta^n_{x^k}(x_1, \ldots, x_n)
=
\left\{
\begin{array}{ll}
\displaystyle
\sum_{i_l \ge 0:i_1+\cdots + i_n=k-n+1}
x_1^{i_1}\cdots x_n^{i_n}
&\hbox{ if } k \ge 0\\
\displaystyle
(-1)^{n-1}
\sum_{i_l \le -1:i_1+\cdots + i_n=k-n+1}
x_1^{i_1}\cdots x_n^{i_n}
&\hbox{ if }k < 0.
\end{array}
\right.\label{4-23}
\end{eqnarray}
In particular, 
when $0 \le k\le n-2$, 
\begin{eqnarray}
\Delta^n_{x^k}(x_1, \ldots, x_n)=0,
\end{eqnarray}
which can be also checked by Lemma \ref{th-23}.
When $k=-1$,
we obtain 
\begin{eqnarray}
\Delta^n_{\frac{1}{x}}(x_1, \ldots, x_n)
&=\frac{(-1)^{n-1}}{x_1\cdots x_n}
\label{4-19-5}.
\end{eqnarray}
For example, 
when $f(x)= \frac{e^x- e^{(1-\alpha)x} - \alpha x}{x^2}$,
the relation (\ref{4-19-5}) yields that
\begin{eqnarray}
\epsilon^j_\alpha(\mu_1, \ldots, \mu_j)
=
(-1)^{j-1} \mu_1 \cdots \mu_j
\Delta^j_{f}(\mu_1, \ldots, \mu_j)
.\label{4-18-7}
\end{eqnarray}
Since the $j$-th derivative of $x^n$ is $0$ for $n \le j-1$,
the relation (\ref{4-18-7}) holds for
$f(x)= \sum_{m=j+1}^{\infty}
(1-(1-\alpha)^m)\frac{x^{m-2}}{m!}$.

Using these formulas, 
we obtain interesting characterization of 
the following two $k \times k$ matrixes $A_k$ and $B_k$:
\begin{eqnarray*}
A_{k:i}^{l}&:=
\left\{
\begin{array}{ll}
\prod_{t=1}^{l-1}(\mu_i-\mu_t)
& \hbox{ if }1 \le l \le i\\
0 & \hbox{ if } i+1\le l 
\end{array}
\right.\\
B_{k:l}^{i}&:=
\left\{
\begin{array}{ll}
\frac{1}{\prod_{t=1, t\neq i}^{l}(\mu_i-\mu_t)}
& \hbox{ if }l \ge i\\
0 & \hbox{ if } 1\le l \le i-1.
\end{array}
\right.
\end{eqnarray*}
That is, 
\begin{eqnarray*}
\fl A_k =
\left(
\begin{array}{cccccc}
A_k^1 & A_k^2 & A_k^3 & \cdots & A_k^{k}
\end{array}
\right)\\
\fl =
\left(
\begin{array}{ccccc}
1 & 0           & 0 & \cdots & 0 \\
1 & \mu_2-\mu_1 & 0 & \cdots & 0 \\
1 & \mu_2-\mu_1 & \prod_{t=1}^2(\mu_3-\mu_t) & \cdots & 0 \\
\vdots & \vdots & \vdots & \ddots & \vdots \\
1 & \mu_2-\mu_1 & \prod_{t=1}^2(\mu_3-\mu_t) & \cdots & 
\prod_{t-1}^{k-1}(\mu_k-\mu_t) 
\end{array}
\right)\\
\fl B_k =
\left(
\begin{array}{c}
B_{k:1}\\
B_{k:2}\\
B_{k:3}\\
\vdots \\
B_{k:k}
\end{array}
\right)\\
\fl =\left(
\begin{array}{ccccc}
1 & 0           & 0 & \cdots & 0 \\
\frac{1}{\mu_1-\mu_2} & \frac{1}{\mu_2-\mu_1} & 0 & \cdots & 0 \\
\frac{1}{\prod_{t=1,\neq 1}^3 (\mu_1-\mu_t)}& 
\frac{1}{\prod_{t=1,\neq 2}^3 (\mu_2-\mu_t)}& 
\frac{1}{\prod_{t=1,\neq 3}^3 (\mu_3-\mu_t)}& \cdots & 0 \\
\vdots & \vdots & \vdots & \ddots & \vdots \\
\frac{1}{\prod_{t=1,\neq 1}^k (\mu_1-\mu_t)}& 
\frac{1}{\prod_{t=1,\neq 2}^k (\mu_2-\mu_t)}& 
\frac{1}{\prod_{t=1,\neq 3}^k (\mu_3-\mu_t)}& \cdots & 
\frac{1}{\prod_{t=1,\neq k}^k (\mu_k-\mu_t)}
\end{array}
\right),
\end{eqnarray*}
where
$A_k^i$ is the $i$-th column vector of the matrix $A_k$
and 
$B_{k:i}$ is the $i$-th row vector of the matrix $B_k$.

These matrixes play an important role in
the proof of Theorems \ref{25-1} and \ref{th-26},
and are the inverse matrixes of each other, i.e.,
\begin{eqnarray}
\sum_{i=1}^k
B_{k:l}^i A_{k:i}^{l'}
= 
\sum_{i=l'}^l
\frac{
\prod_{t=1}^{l'-1}(\mu_i-\mu_t)
}{\prod_{t=1, t\neq i}^{l}(\mu_i-\mu_t)}=
\delta^{l,l'},\label{4-19-1}
\end{eqnarray}
which is equivalent with 
\begin{eqnarray}
\sum_{l=1}^k
B_{k:l}^i A_{k:i'}^{l}
=\delta_{i,i'}.\label{4-19-1-2}
\end{eqnarray}
The equation (\ref{4-19-1}) is trivial in the case of $l'\ge l$.
When $l' < l$,
\begin{eqnarray}
\sum_{i=l'}^l
\frac{
\prod_{t=1}^{l'-1}(\mu_i-\mu_t)
}{\prod_{t=1, t\neq i}^{l}(\mu_i-\mu_t)}=
\sum_{i=l'}^l
\frac{1}{\prod_{t=l', t\neq i}^{l}(\mu_i-\mu_t)}
\label{4-19-2}.
\end{eqnarray}
Applying Lemma \ref{th-23} to the case of $n=l-l'+1$ and
$f(x)= 1$,
we can show that LHS of (\ref{4-19-2}) is equal to $0$.

Now, we modify the matrix $A_k$ as follows:
\begin{eqnarray*}
C_{k:i}^l:=
\left\{
\begin{array}{ll}
\frac{1}{\mu_i}
&\hbox{ if }l=1\\
\prod_{t=1}^{l-2}(\mu_i-\mu_t)
& \hbox{ if }2 \le l \le i+1\\
0 & \hbox{ if } i+2\le l .
\end{array}
\right.
\end{eqnarray*}
That is,
\begin{eqnarray*}
C_k &=
\left(
\begin{array}{cccccc}
\frac{1}{\mu_1} & 1 & 0           & 0 & \cdots & 0 \\
\frac{1}{\mu_2} & 1 & \mu_2-\mu_1 & 0 & \cdots & 0 \\
\frac{1}{\mu_3} & 1 & \mu_2-\mu_1 & \prod_{t=1}^2(\mu_3-\mu_t) & \cdots & 0 \\
\vdots & \vdots & \vdots & \vdots & \ddots & \vdots \\
\frac{1}{\mu_{k-1}} & 1 & \mu_2-\mu_1 & \prod_{t=1}^2(\mu_3-\mu_t) & \cdots & 
\prod_{t-1}^{k-1}(\mu_k-\mu_t) \\
\frac{1}{\mu_{k}} & 1 & \mu_2-\mu_1 & \prod_{t=1}^2(\mu_3-\mu_t) & \cdots & 
\prod_{t-1}^{k-1}(\mu_k-\mu_t) 
\end{array}
\right)\\
&=
\left(
\begin{array}{cccccc}
A_k^0 & A_k^1 & A_k^2 & A_k^3 & \cdots & A_k^{k-1}
\end{array}
\right),
\end{eqnarray*}
where
\begin{eqnarray*}
A_k^0= \left(
\begin{array}{c}
\frac{1}{\mu_1} \\
\frac{1}{\mu_2} \\
\frac{1}{\mu_3} \\
\vdots \\
\frac{1}{\mu_{k-1}} \\
\frac{1}{\mu_{k}} 
\end{array}
\right).
\end{eqnarray*}
The inverse matrix $C_k^{-1}$ is characterized as follows:
\begin{eqnarray}
C_k^{-1}=
\left(
\begin{array}{c}
\vec{0}^T\\
B_{k:1}\\
B_{k:2}\\
B_{k:3}\\
\vdots \\
B_{k:k-1}
\end{array}
\right)
-(-1)^k
\left(
\begin{array}{cc}
\mu_1\mu_2\cdots\mu_k B_{k:k}\\
-\mu_2\mu_3\cdots\mu_k B_{k:k}\\
\mu_3\mu_4\cdots\mu_k B_{k:k}\\
-\mu_4\mu_5\cdots\mu_k B_{k:k}\\
\vdots \\
(-1)^{k-1}\mu_k B_{k:k}
\end{array}
\right).\label{4-27-1}
\end{eqnarray}
The equation (\ref{4-27-1}) can be checked as follows.
Using (\ref{4-19-5}), we obtain
\begin{eqnarray*}
\mu_1\mu_2\cdots\mu_k B_{k:k}A_k^0=\sum_{i=1}^k
(-1)^{k-1}
\frac{\mu_1\cdots \mu_k}
{\prod_{t=1,t\neq i}^{k}(\mu_i-\mu_t)}
C_{k:i}^1
=1,
\end{eqnarray*}

Since
\begin{eqnarray*}
B_{k:i-1}
A_k^0
=
\sum_{l=1}^{i-1}\frac{1}
{\mu_l\prod_{t=1,t\neq l}^{i-1}(\mu_l-\mu_t)}
=
\frac{(-1)^{i-2}}{\mu_1 \mu_2 \cdots \mu_{i-1}}\\
(-1)^{k+i}\mu_i\mu_{i+1}\cdots\mu_k B_{k:k}
A_k^0
= (-1)^{k+i}\mu_i\mu_{i+1}\cdots\mu_k
\frac{(-1)^{k-1}}{\mu_1\cdots \mu_k},
\end{eqnarray*}
we have
\begin{eqnarray*}
(B_{k:i-1}+
(-1)^{k+i}\mu_i\mu_{i+1}\cdots\mu_k B_{k:k})
A_k^0=0
\end{eqnarray*}
for $k \ge i \ge 2$.
Further
\begin{eqnarray*}
(B_{k:i-1}+
(-1)^{k+i}\mu_i\mu_{i+1}\cdots\mu_k B_{k:k})
A_k^j=B_{k:i-1} A_k^{j-1}= \delta_i^j
\end{eqnarray*}
for $k \ge i \ge 2, k \ge j \ge 2$,
and 
\begin{eqnarray*}
-(-1)^{k}\mu_1\mu_{2}\cdots\mu_k B_{k:k})A_k^{j-1}=0
\end{eqnarray*}
for $k \ge j \ge 2$.
Thus, we obtain (\ref{4-27-1}).

\subsection{Proof of Theorem \ref{25-1}}
Now, we prove Theorem \ref{25-1}.
First, we check the positivity of 
$\rho_i$, which is equivalent with
the positivity of $\gamma_{l,n}$ for $i= 2, \ldots, k+1$,
$n \ge 2$.
Substituting 
$l-2$ and $x^{n-2}$
into $n$ and $f(x)$
in Theorem \ref{th-23}, we have
\begin{eqnarray*}
&{(l-2)!}
\gamma_{l,n}
=
{(l-2)!}
\sum_{j=1}^{l-1}
\frac
{\mu_j^{n-2}}
{\prod_{t=1,t\neq j}^{l-1}(\mu_j-\mu_{t})}
\\
=&
\int_{\Delta_{l-2}}\!
\frac{(n-2)!}{(n-l)!}( \sum_{t=1}^{l-1} a_t \mu_t )^{n-l}
p_{l-2} (a_1,\!\ldots\!, a_{l-1})
d a_1\!\ldots \!d a_{l-1}
\ge 0.
\end{eqnarray*}
The equation (\ref{10-20-2}) is equivalent with 
\begin{eqnarray*}
\mu_s^2\sum_{n=2}^{\infty}\frac{\mu_s^{n-2}}{n !}|n \rangle \langle n|
=
\mu_s^2 \sum_{l=2}^{s+1}
\prod_{t=1}^{l-2}
(\mu_i-\mu_t)
\Omega_l 
\rho_l\\
\end{eqnarray*}
Applying (\ref{4-19-1-2}) to the case of $k=s$,
any function $f$ satisfies 
\begin{eqnarray*}
f(\mu_s)=\sum_{l=1}^s\sum_{j=1}^{l}
\frac{\prod_{t=1}^{l-1}(\mu_s-\mu_t)}
{\prod_{t=1,t\neq j}^{l}(\mu_j-\mu_t)}
f(\mu_j).
\end{eqnarray*}
When $f(x)=x^{n-2}$ and $n-2 \le l-2$,
$f^{(l-1)}(x)=0$.
Thus, Theorem \ref{th-23} yields that
\begin{eqnarray*}
\sum_{j=1}^{l}
\frac{1}
{\prod_{t=1,t\neq j}^{l}(\mu_j-\mu_t)}
f(\mu_j)=0.
\end{eqnarray*}
Therefore,
\begin{eqnarray}
&\sum_{n =2}^{\infty}
\frac{\mu_s^{n-2}}{n!}|n \rangle \langle n|
=
\sum_{n =2}^{\infty}
\sum_{l=1}^{s}
\sum_{j=1}^{l}
\frac{\prod_{t=1}^{l-1}(\mu_s-\mu_t)}
{\prod_{t=1,t\neq j}^{l}(\mu_j-\mu_t)}
\frac{\mu_j^{n-2}}{n!}|n \rangle \langle n|\nonumber \\
=&
\sum_{l=1}^{s}
\prod_{t=1}^{l-1}(\mu_s-\mu_t)
\sum_{n =l+1}^{\infty}
\sum_{j=1}^{l}
\frac{\mu_j^{n-2}}
{n!\prod_{t=1,t\neq j}^{l}(\mu_j-\mu_t)}
|n \rangle \langle n|\nonumber \\
=&
\sum_{l=1}^{s}
\prod_{t=1}^{l-1}(\mu_s-\mu_t)
\sum_{n =l+1}^{\infty}
\frac{\gamma_{l+1,n}}
{n!}
|n \rangle \langle n|
=
\sum_{l=1}^{s}
\prod_{t=1}^{l-1}
(\mu_i-\mu_t)
\Omega_{l+1} 
\rho_{l+1}\nonumber ,
\end{eqnarray}
which implies (\ref{10-20-2}).

\subsection{Proof of Theorem \ref{th-26}}
Next, we prove Theorem \ref{th-26}.
When we fix the parameter $q^{1+j}$ as well as $q^0$,
the parameters $q^1, \ldots , q^{j}$ are uniquely decided only from 
the parameters $p_1, \ldots, p_j$.
That is, the relation
\begin{eqnarray*}
\tilde{q}_{j,\min}&:=
\min_{\vec{q}=(q^0,\ldots,q^{2k+1})^T}
\left\{q^1\left|
\begin{array}{l}
p_i = \sum_{j=0}^{2k+1} P_i^j q^j+p_D\\
1-p_D\ge q^{1+j}\ge 0
\end{array}
\right.\right\}\\
&= \min_{\vec{q}=(q^0,\ldots,q^{j+1})^T}
\left\{q^1\left|
\begin{array}{l}
p_i = \sum_{t=0}^{j+1} P_i^j q^t+p_D\\
\hbox{ for }i=1, \ldots, j\\
1-p_D\ge q^{1+j}\ge 0
\end{array}
\right.\right\}.
\end{eqnarray*}
Now, we focus on 
parameters $q^1, \ldots, q^{j+1}$.
Then, we have
\begin{eqnarray*}
&p_i-p_D -e^{-\mu_i}(p_0-p_D)\\
=& \sum_{s=1}^{j} 
e^{-\mu_i}\mu_i^2 
C_{j:i}^{s}\Omega_s q^{s}
 +
e^{-\mu_j}\mu_j^2 \prod_{t=1}^{j-1}(\mu_j-\mu_t)
\Omega_{j+1}q^{j+1} \delta_{i,j}
\end{eqnarray*}
for $i=1,\ldots, j$.
Using (\ref{4-27-1}) and (\ref{4-27-2}), we have
\begin{eqnarray*}
q^1
&=
\sum_{i=1}^j
(C_j^{-1})_1^i
\frac{e^{\mu_i}}{\mu_i^2}
\Bigl(p_i-p_D
-e^{-\mu_i}(p_0-p_D)
-
e^{-\mu_j}\mu_j^2 \prod_{t=1}^{j-1}(\mu_j-\mu_t)
\Omega_{j+1}q^{j+1}\delta_{i,j}
\Bigr)\\
&=
\sum_{i=1}^j
\beta_i^j
\Bigl(p_i-p_D
-e^{-\mu_i}(p_0-p_D)\Bigr)
-(-1)^{j-1}
\mu_1 \cdots \mu_j\Omega_{j+1}
q^{j+1}.
\end{eqnarray*}
Note that $\Omega_{j+1}\ge 0$.
Since $0 \le q^{j+1}\le 1-p_D$,
we obtain (\ref{11-12-1}).
Similarly, we can prove 
(\ref{11-12-2}) and (\ref{11-12-8}).

\subsection{Proof of Theorem \ref{th-26-2-a}}
Choose $f$ as $f(x)=\frac{e^x-e^{(1-\alpha)x}-\alpha x}{x^2}
= \sum_{n=1}\frac{1}{n!}(1-(1-\alpha)^n)x^{n-2}$.
Using (\ref{4-23}) and Lemma \ref{th-23},
we obtain
\begin{eqnarray*}
\fl ~\mu_1 \cdots \mu_j
\sum_{n=j+1}^{\infty}
\sum_{i_l\ge0:i_1+\cdots+i_j=n-1-j}
\frac{1}{n!}(1-(1-\alpha)^n)
\mu_1^{i_1}\cdots\mu_1^{i_j}
\\
\fl=\mu_1 \cdots \mu_j\Delta_f(\mu_1, \ldots, \mu_j)\\
\fl=\mu_1\cdots \mu_j
\sum_{i=1}^j
\frac{ \sum_{n=1}^{\infty}\frac{1}{n!}
(\mu_i^n-((1-\alpha)\mu_i)^n)}
{\mu_i^2 
\prod_{1\le k \le j, k\neq i}(\mu_i-\mu_k)}\\
\fl=
\frac{\mu_1\cdots\mu_j}{(j-1)!}
\int_{\Delta_{j-1}}
\sum_{n=0}^\infty
\frac{1-(1-\alpha)^{n+j+1}}{(n+j+1)(n+j)n!}
(\sum_{i=1}^{j}a_i \mu_i)^{n}
p_{j-1}(a_1\ldots a_{j})d a_1\ldots d a_{j}.
\end{eqnarray*}
Thus, the relations (\ref{11-4-1}) and (\ref{4-23-2}) 
follow from (\ref{4-18-7}).

Apply (\ref{4-18-7}) to the case of
$f(x)= \sum_{m=j+1}^{\infty}\frac{x^{n-2}}{m!}$.
We obtain
\begin{eqnarray*}
&\mu_1 \cdots \mu_j
\Omega_{j+1}
=
\mu_1 \cdots \mu_j
\left(\sum_{s=1}^{j}
\frac{
\sum_{n=j+1}^{\infty}\frac{\mu_s^{n-2}}{n!}
}{\prod_{t=1,t\neq s}^{j}(\mu_s-\mu_t)}\right)
=
\epsilon_{1}^j(\mu_1,\ldots,\mu_j ),
\end{eqnarray*}
which implies (\ref{10-26-11}).

\section{Conclusion and further improvement}
We have discussed the AKG rates with phase-randomized coherent light
by the decoy method, in which the number $k$ of possible intensities is
arbitrary. For this purpose, 
by generalizing Wang's expansion,
we have derived a convex expansion of the
phase-randomized coherent state, which allows us to parameterize Eve's
operation using $3k+3$ parameters even in the general case. 
Thanks to this parameterization,
lower bound of AKG rate has been obtained
with $k$ decoy intensities. Also, assuming
that the noise in the quantum channel is described by a natural model,
we have derived upper bounds independent of the number $k$ of
decoy intensities as the universal upper bound. 
It has been numerically demonstrated that 
the AKG rate is close to the universal upper bound in the case of $k = 3$,
i.e., the AKG rates cannot be improved so much even if we prepare a
number of intensities $k$ larger than $3$.

Further, we have discussed the intensity maximizing our upper bound of
the AKG rate. It has been shown that this optimum intensity is always
less than $1$. We have also characterized the relationship between the
AKG rate and intensities that do not generate raw keys. These results
indicate how to choose intensities in an implemented quantum key
distribution system, in which the decoy method is applied.
Unfortunately, this paper does not treat the security with the finite-length code.
However, we will treat this issue in the papers \cite{H3,HHHT},
in which our expansion (\ref{10-20-2}) and the matrix $P_k$ play 
an essential role.

For a further improvement of AKG rate,
we can assume that 
some error happens in the generator or the detector.
In this case, 
there is a relation 
${r^1}'= (1-p_S)r^1+p_S(1-r^1)$
between the observed error rate ${r^1}'$ in the $\times$ basis
and 
the error rate ${r^1}'$ out side of the generator and the detector
in the $\times$ basis,
where 
$p_S$ is the probability that the error in the $\times$ basis 
occurs at generation or detection in the single-photon state.
That is, it is suitable to substitute 
$\frac{\frac{b^1_{\max}}{q^1_{\min}}-p_S}{1-2 p_S}$
in side of the binary entropy $h$.
If it is possible to distinguish
the error probabilities at generation and detection,
a further improvement is available.
By taking into account bit error probability
among single-photon states at generating the pulse,
a tighter evaluation of the AKG rate of the forward case may be possible 
in a way similar to Renner et al.\cite{RGK}.

\section*{Acknowledgments}
The author would like to thank Professor Hiroshi Imai of the
ERATO-SORST, QCI project for support.
He is grateful to Professor Hiroshi Imai, Dr. Akihisa Tomita, 
Dr. Tohya Hiroshima, and Mr. Jun Hasewaga for useful discussions.
He thanks Dr. Francesco Buscemi for his helpful comments.

\appendix

\section{Proof of Lemma \ref{th-23}}\label{a1}
We will prove (\ref{10-23-1})
by induction.
The case of $n=1$ is trivial.
By the assumption of induction, we obtain the following equations.
\begin{eqnarray}
\fl \int_{\Delta_{n-1}}
f^{(n-1)}(\sum_{i=1}^{n}a_i x_i)
p_{n-1}(a_1\ldots a_{n})
d a_1\ldots d a_{n}
=(n-1)!
\sum_{i=1}^{n} 
\frac{f(x_i)}{\prod_{1\le j \le n, j \neq i}(x_i-x_j)},
\label{10-23-2}\\
\fl \int_{\Delta_{n-1}}
f^{(n-1)}(\sum_{i=2}^{n+1}a_i x_i)
p_{n-1}(a_2\ldots a_{n+1})
d a_2\ldots d a_{n+1}
=(n-1)!
\sum_{i=2}^{n+1} 
\frac{f(x_i)}{\prod_{2\le j \le n+1, j \neq i}(x_i-x_j)}.\nonumber\\
\label{10-23-3}
\end{eqnarray}
Thus,
\begin{eqnarray*}
\fl
\int_{\Delta_{n-1}}
p_{n-1}(a_1\ldots a_{n})
f^{(n-1)}(\sum_{i=2}^{n+1}a_i x_i)
d a_2\ldots d a_{n+1}\\
\fl \quad -
\int_{\Delta_{n-1}}
f^{(n-1)}(\sum_{i=1}^{n}a_i x_i)
p_{n-1}(a_2\ldots a_{n+1})
d a_1\ldots d a_{n}\\
\fl =
(n-1)!
\Bigl(
\sum_{i=2}^{n}
\frac{f(x_i)
\bigl(
\frac{1}{x_i-x_{n+1}}
-
\frac{1}{x_i-x_{1}}
\bigr)
}{\prod_{2\le j \le n, j \neq i}(x_i-x_j)}
+\frac{f(x_1)}{\prod_{2\le j \le n}(x_1-x_j)}
-\frac{f(x_{n+1})}{\prod_{2\le j \le n, }(x_{n+1}-x_j)}
\Bigr)\\
\fl =
(n-1)!
\Bigl(
\sum_{i=2}^{n}
\frac{f(x_i)}{\prod_{2\le j \le n, j \neq i}(x_i-x_j)}
\frac{x_{n+1}-x_1  }{(x_i-x_{n+1})(x_i-x_{1})}\\
\fl \qquad -\frac{f(x_1)}{\prod_{2\le j \le n}(x_1-x_j)}
+\frac{f(x_{n+1})}{\prod_{2\le j \le n, }(x_{n+1}-x_j)}
\Bigr)\\
\fl =
(n-1)!(x_{n+1}-x_1)
\Bigl(
\sum_{i=2}^{n}
\frac{f(x_i)}{\prod_{1\le j \le n+1, j \neq i}(x_i-x_j)}
+\frac{f(x_1)}{\prod_{2\le j \le n+1}(x_1-x_j)}
+\frac{f(x_{n+1})}{\prod_{1\le j \le n, }(x_{n+1}-x_j)}
\Bigr)\\
\fl =
(n-1)!(x_{n+1}-x_1)
\sum_{i=1}^{n+1} 
\frac{f(x_i)}{\prod_{1\le j \le n+1, j \neq i}(x_i-x_j)}.
\end{eqnarray*}
Now, we introduce new parameters
$b= 1-a_1$ or $1-a_{n+1}$ and 
$b_i= \frac{a_i}{b}$ $(i=2,\ldots,n)$.
Then,
\begin{eqnarray}
\fl ~(n-1)!(x_{n+1}-x_1)
\sum_{i=1}^{n+1} 
\frac{f(x_i)}{\prod_{1\le j \le n+1, j \neq i}(x_i-x_j)}\nonumber\\
\fl =
(n-1)
\int_{\Delta_{n-2}}
\int_{0}^1
f^{(n-1)}(\sum_{i=2}^{n}b b_i x_i + (1-b) x_{n+1})
d b ~b^{n-2} p_{n-2}(b_2\ldots b_{n})d b_2\ldots d b_{n} \nonumber\\
\fl \quad -
(n-1)\int_{\Delta_{n-2}}
\int_{0}^1
f^{(n-1)}(\sum_{i=2}^{n}b b_i x_i+ (1-b)x_1)
d b ~b^{n-2} p_{n-2}(b_2\ldots b_{n})d b_2\ldots d b_{n} \nonumber\\
\fl =
(n-1)
\int_{\Delta_{n-2}}
\int_{0}^1
\left(f^{(n-1)}(\sum_{i=2}^{n}b b_i x_i + (1-b) x_{n+1})
-f^{(n-1)}(\sum_{i=2}^{n}b b_i x_i+ (1-b)x_1)
\right)\nonumber\\
\fl \quad d b ~b^{n-2} p_{n-2}(b_2\ldots b_{n})d b_2\ldots d b_{n}\nonumber \\
\fl =
(n-1)
(x_{n+1}-x_1)
\int_{\Delta_{n-2}}
\int_{0}^1
\int_{0}^{1}
f^{(n)}(\sum_{i=2}^{n}b b_i x_i + 
(1-b) (x_{n+1}-x_1)s
+ (1-b)x_1
)
d s \nonumber\\
\fl \quad d b ~(1-b) b^{n-2} p_{n-2}(b_2\ldots b_{n})d b_2\ldots d b_{n} \nonumber\\
\fl =
\frac{1}{n}
(x_{n+1}-x_1)
\int_{\Delta_{n}}
f^{(n)}(\sum_{i=1}^{n+1} c_i x_i )
p_{n}(b_1\ldots b_{n+1})
d c_1 \ldots d c_{n+1} 
\label{10-23-6}
\end{eqnarray}
In equation (\ref{10-23-6}), we introduce parameters
$c_1 = (1-b)(1-s)$
$c_i=  b b_i$  $(i=2, \ldots, n)$,
$c_{n+1} = (1-b)s$,
and use the relation
$(1-b) d c_1 d c_{n+1} =ds db$.
The coefficient $\frac{1}{n}$ in (\ref{10-23-6})
can be checked by the relation
\begin{eqnarray*}
\int_{\Delta_{n-2}}
\int_{0}^1
\int_{0}^{1}
ds db (1-b)b^{n-2} p_{n-2}(b_2\ldots b_{n})d b_2\ldots d b_{n} 
= \frac{1}{n(n-1)}.
\end{eqnarray*}
Therefore, we obtain (\ref{10-23-1}).

\section{Proof of Lemma \ref{lem1}}\label{a2}
Since 
$\frac{1}{(x_i-x_1)(x_i-x_{n+1})}
=
(\frac{1}{x_i-x_{n+1}}
-\frac{1}{x_i-x_1})
\frac{1}{x_{n+1}-x_1}$,
$\Delta^{n+1}_f(x_1, \ldots, x_{n+1})$ is calculated as follows:
\begin{eqnarray*}
\fl ~ \Delta^{n+1}_f(x_1, \ldots, x_{n+1})\\
\fl =
\frac{1}{(x_i-x_1)(x_i-x_{n+1})}
\sum_{i=2}^{n} 
\frac{f(x_i)}{\prod_{j=2,j \neq i}^{n}(x_i-x_j)}
+
\frac{f(x_1)}{\prod_{j=2}^{n+1}(x_1-x_j)}
+
\frac{f(x_{n+1})}{\prod_{j=1}^{n}(x_{n+1}-x_j)}\\
\fl =
\frac{1}{x_{n+1}-x_1}
\sum_{i=2}^{n} 
\frac{f(x_i)}{\prod_{j=2,j \neq i}^{n+1}(x_i-x_j)}
-
\frac{1}{x_{n+1}-x_1}
\sum_{i=2}^{n} 
\frac{f(x_i)}{\prod_{j=1,j \neq i}^{n}(x_i-x_j)}
\\
\fl \quad +
\frac{1}{x_1-x_{n+1}}
\frac{f(x_1)}{\prod_{j=2}^{n}(x_1-x_j)}
+
\frac{1}{x_{n+1}-x_{1}}
\frac{f(x_{n+1})}{\prod_{j=2}^{n}(x_{n+1}-x_j)}\\
\fl =
\frac{1}{x_{n+1}-x_1}
\Delta^{n}_f(x_2, \ldots, x_{n+1})
-\frac{1}{x_{n+1}-x_1}
\Delta^{n}_f(x_1, \ldots, x_{n}).
\end{eqnarray*}

\section{Proof of (\ref{11-5-10}) and (\ref{11-5-11})}\label{4-13-1}
It is sufficient to show
\begin{eqnarray}
\frac{d I_{\to}(\mu;p_0,p_D,p_S,\alpha,s)}{d \mu}|_{\mu=1}&\le 0
\label{11-5-30}\\
\frac{d I_{\leftarrow}(\mu;p_0,p_D,p_S,\alpha,s)}{d \mu}|_{\mu=1}&\le 0.
\label{11-5-31}
\end{eqnarray}
From the assumption, the parameter $s'$ also satisfies 
$0 \le s'\le 1/2$.
Hence, 
$\overline{h}
(\frac{s'(1-e^{-\alpha \mu})+\frac{1}{2}p_0}{1-e^{-\alpha \mu}+p_0})$
is monotone increasing concerning $\mu$,
i.e., 
$\frac{d \overline{h}
(\frac{s'(1-e^{-\alpha \mu})+\frac{1}{2}p_0}{1-e^{-\alpha \mu}+p_0}))}
{d \mu}\ge 0$.
Since
\begin{eqnarray*}
&\frac{d I_{\to}(\mu;p_0,p_D,p_S,\alpha,s)}{d \mu}\\
=&
\frac{1}{2}\Bigl((1-\mu) e^{-\mu}
(\alpha +(p_0-p_D))
(1-\overline{h}(\frac{s\alpha +\frac{1}{2}(p_0-p_D)}{\alpha +(p_0-p_D)}))\\
&- e^{-\mu}p_0
- (1-e^{-\alpha \mu}+p_0)
\frac{d \overline{h}
(\frac{s'(1-e^{-\alpha \mu})+\frac{1}{2}p_0}{1-e^{-\alpha \mu}+p_0})}
{d \mu} 
- e^{-\alpha \mu}
\overline{h}
(\frac{s'(1-e^{-\alpha \mu})+\frac{1}{2}p_0}{1-e^{-\alpha \mu}+p_0})
\Bigr),\\
&\frac{d I_{\leftarrow}(\mu;p_0,p_D,p_S,\alpha,s)}{d \mu}\\
=&
\frac{1}{2}\Bigl((1-\mu) e^{-\mu}
(\alpha +(p_0-p_D))
(1-\overline{h}(\frac{s\alpha +\frac{1}{2}(p_0-p_D)}{\alpha +(p_0-p_D)}))
\\
&- (1-e^{-\alpha \mu}+p_0)
\frac{d \overline{h}
(\frac{s'(1-e^{-\alpha \mu})+\frac{1}{2}p_0}{1-e^{-\alpha \mu}+p_0})}
{d \mu} 
- e^{-\alpha \mu}
\overline{h}
(\frac{s'(1-e^{-\alpha \mu})+\frac{1}{2}p_0}{1-e^{-\alpha \mu}+p_0})
\Bigr),
\end{eqnarray*}
we can check (\ref{11-5-30}) and (\ref{11-5-31}).

\end{document}